\documentclass[twocolumn,preprintnumbers,superscriptaddress,aps,prl,amsmath,amssymb]{revtex4}
\usepackage{graphicx} % Include figure files
\usepackage{dcolumn}  % Align table columns on decimal point
\usepackage{color}
\usepackage{hyperref}
\usepackage{breakurl}

\newcommand{\SiO}{SiO$_2$}

\begin{document}

\title{Si nanoparticle interfaces in Si/\SiO\ solar cell materials}
\author{S. Kilpel\"ainen}
\affiliation{Department of Applied Physics, Aalto University,
P.O.Box 11100 FI-00076 Aalto, Finland}

\author{Y.-W. Lu}
\affiliation{Department of Physics and Astronomy, Aarhus University,
DK-8000 Aarhus C, Denmark}

\author{F. Tuomisto}
\affiliation{Department of Applied Physics, Aalto University,
P.O.Box 11100 FI-00076 Aalto, Finland}

\author{J. Slotte}
\affiliation{Department of Applied Physics, Aalto University,
P.O.Box 11100 FI-00076 Aalto, Finland}

\author{A. Nylandsted Larsen}
\affiliation{Department of Physics and Astronomy, Aarhus University,
DK-8000 Aarhus C, Denmark}

\date{\today}

\begin{abstract}
Novel solar cell materials consisting of Si nanoparticles embedded
in \SiO\ layers have been studied using positron annihilation
spectroscopy in Doppler broadening mode, and photoluminescence. Two
positron-trapping interface states are observed after high
temperature annealing at 1100$^{\circ}$C. One of the states is
attributed to the (\SiO/Si bulk) interface and the other to the
interface between the Si nanoparticles and \SiO. A small reduction
in positron trapping into these states is observed after annealing
the samples in N$_2$ atmosphere with 5\% H$_2$. Enhanced
photoluminescence is also observed from the samples following this
annealing step.
\end{abstract}

\maketitle

Solar cell industry has relied on silicon as the mainstream material
since the discovery of the Si-based photocell~\cite{Chapin1954}. The
advantages of Si solar cells are unambiguous; they are easy to
produce in masses, relatively cheap, can be used in various
applications, and Si is non-toxic. However, the energy conversion
efficiency is just approaching 25\% for the very best commercial Si
solar cells, and the typical mainstream cells have efficiencies of
around 15\% ~\cite{Gizmag2010}. The classical theoretical limit for
single Si solar cells, based on the energy spectrum of the Sun and
the band gap of Si, is not much better: 31\%~\cite{Marti1996}. A
typical way to produce very efficient solar cells is to use
composite structures consisting of layers made from different
materials sensitive to different wavelengths of
sunlight~\cite{Fraas1981}. These multilayered solar cell structures
are, however, very complex to produce and thus expensive so they are
not suitable for mainstream use. Since traditional energy sources
such as coal and oil are gradually running out and on top of that
pollute the Earth's atmosphere, researchers are very eager to find
ways to better harness the energy of the Sun. A promising concept
was discovered a couple of years ago when Klimov and his colleagues
discovered that silicon nanocrystals could convert a single photon
into more than one excitons~\cite{Schaller2004}. A solar cell
utilizing this property could be a workaround for getting beyond
silicon's theoretical energy conversion efficiency without having to
resort to expensive materials. Promising candidates for such solar
cells could be silica layers embedded with Si nanocrystals. Such
layers have been prepared and studied with several
techniques~\cite{Cho2004,Cho2004a,Conibeer2006,Hao2009,Osinniy2009,Zacharias2002,Surana2010}.
The results have been mostly promising but there is still much work
to be done before these solar cells can become commercially
available. The biggest problem is that the interface between the
nanocrystals and silica tends to have carrier traps in it, thus
decreasing the light conversion efficiency.

Positron annihilation spectroscopy (PAS) is a versatile tool for
studying vacancy-type defects in various
materials.~\cite{Krause-Rehberg,Saarinen1998} The annihilation
radiation provides atomic resolution which is useful in getting
information about the defects themselves as well as about the atoms
in their near vicinity. The use of slow, monoenergetic positrons
allows the study of defect distributions in thin layers and
interfaces.

In this work, we used a monoenergetic slow positron beam in Doppler
broadening mode to study interfaces between silicon nanocrystals and
\SiO\ which they were embedded in. The fast positrons emitted by a
$^{22}$Na source were moderated with a 1 $\mu$m tungsten foil,
accelerated with an electric field and then implanted into the
sample at energies ranging from 0.5 to 35 keV. After implantation,
the positron thermalizes rapidly in the sample; the thermalization
time is but a few picoseconds. Following thermalization, the
positron diffuses in the sample for ~100-250 ps (tens to hundreds of
nanometers depending on the material and the defects present) before
it finds an electron and
annihilates.~\cite{Saarinen1998,Krause-Rehberg} Positrons can get
trapped into neutral and negatively charged open volume defects in
the sample. Particularly, open volume defects such as vacancies act
as efficient positron traps. Such defects can be characterized by
increased positron lifetime in lifetime measurements and a narrower
momentum distribution in Doppler broadening measurements.

In positron Doppler broadening spectroscopy, the broadening of the
511 keV annihilation line due to the momentum of the annihilating
electron-positron pair is detected. In this work, two high purity Ge
detectors with an energy resolution of ~1.2 keV at the 511 keV line
were used to detect the annihilation quanta and to measure their
energies. When a positron annihilates with an electron, two 511 keV
annihilation quanta are emitted at almost the opposite directions,
with a small angular difference induced as an additional effect of
the momentum of the annihilating pair being conserved in the
process.

In the standard Doppler broadening measurements, the results are
typically described with the conventional line shape parameters $S$
and $W$. The $S$ parameter, also often referred to as the low
momentum parameter, is defined as the fraction of counts in the
central part of the annihilation peak. Annihilations contributing to
this part of the spectrum correspond mainly to valence electrons.
The high momentum parameter $W$, analogously, tells the fraction of
counts in both wings of the annihilation peak and corresponds mainly
to annihilations with core electrons. The energy windows for both
parameters are typically chosen so that the sensitivity of both is
at maximum when it comes to changes in the annihilation environment.
In this work, the windows were set to $|$p$_z$$| <$ 0.44 a.u. for
$S$ and 1.60 a.u. $< |$p$_z$$| <$ 4.10 a.u. for $W$. Open volume
defects have a reduced electron density and this narrows the
Doppler-broadened spectrum. Thus, an elevated $S$ (or lowered $W$)
parameter typically indicates the presence of open volume defects in
a sample.

The measured line shape parameters $S$ and $W$ are always
superpositions of the $S$ and $W$ parameters of different positron
states in the sample. In the simplest case, there are only two
possible annihilation states (surface and bulk, or bulk and a
defect) and the measured parameters can be obtained from

\begin{equation}
S=\eta_1S_1+\eta_2S_2
\end{equation}
\begin{equation}
W=\eta_1W_1+\eta_2W_2,
\end{equation}

\noindent where $S_i$ $(W_i)$ is the $S$ $(W)$ parameter of state
$i$ and $\eta_i$ the annihilation fraction in state $i$. The
equations above being parameterized equations of lines in the ($S$,
$W$) plane is very useful; plotting the measurement results in the
($S$, $W$) plane and analyzing the slopes of the aforementioned
lines helps in identifying the defects present in the sample. Also,
any nonlinear behavior in an ($S$, $W$) plot is an indication of
three or more positron annihilation states. These can be any
combination of surface, bulk, defect and other positron trapping
states such as interfaces.

Multilayer structures consisting of 30 Si/\SiO\ bilayers capped by
an SiO$_2$ layer of 50 nm were deposited on oriented p-type Si (100)
substrates by using an RF magnetron sputtering system without
substrate heating. Si and \SiO\ targets were alternately sputtered
in Ar gas at 3 mTorr. In the bilayers, the thickness of each \SiO\
layer was 4 nm while Si layers of thicknesses of 1, 2 and 4 nm were
employed in the three samples named as the 1 nm, 2 nm and 4 nm
samples, respectively. Thus, the total thickness of the layer
structure was either 200 nm, 230 nm or 290 nm. A pure \SiO\ layer of
300 nm used as a reference was also deposited on p-type Si substrate
by magnetron sputtering without substrate heating.

In order to form the Si nanocrystals, all samples were annealed in
N$_2$ at 1100$^{\circ}$C for one hour. Then each annealed sample was
cut into two pieces and one of them was further annealed in 95\%
N$_2$ + 5\% H$_2$ at 500$^{\circ}$C for one hour in order to
passivate defects~\cite{Wilkinson2003,Yedji2011,Koponen2009}.
Positron measurements were then performed on all three types of
samples (as-deposited, once annealed and twice annealed). All
measurements were done with a slow positron beam in Doppler
broadening mode at room temperature. Photoluminescence measurements
were also performed on all samples. A continuous wave laser (Oxxius
Violet) with a wavelength of 405 nm and excitation power of 50 mW
was used as the excitation source. The PL spectra were measured at
room temperature using a single monochromator (dispersion 0.8 nm/mm,
resolution 0.008 nm) and a silicon photodiode. All spectra were
corrected for the spectral response of the detection system.

In Fig. 1, the positron $S$ parameter is shown as a function of
positron implantation energy for the as-deposited and the two
annealed samples with 2 nm Si layers. The layer structure is easily
seen in the as-deposited sample as a flat plateau at implantation
energies 1-5 keV. This region looks completely different in the
sample annealed at 1100$^{\circ}$C. There is a sharp peak at $\sim$2
keV and a valley at around 5 keV. Both of these features indicate a
decrease in the positron diffusion length L$_+$ which means that
positrons get trapped either at the interfaces or in interfacial
defects. The second annealing step at 500$^{\circ}$C involving H
brings the $S$ parameter down which is a sign of reduced positron
trapping.

\begin{figure}[t]
\centering{\includegraphics[angle=0, width=9cm]{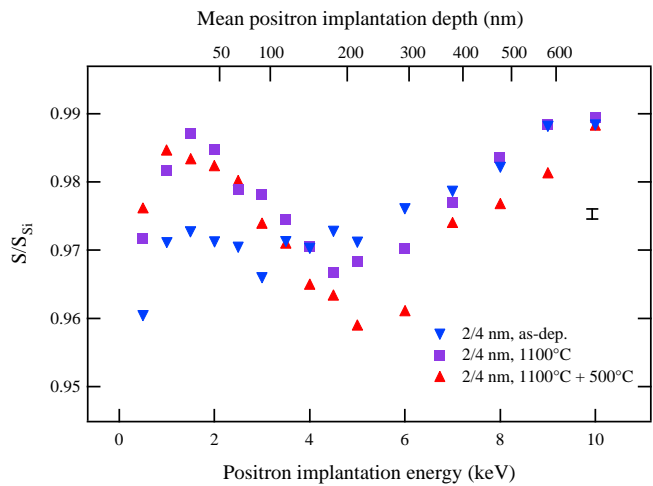}}
\caption{The $S$ parameters as a function of positron implantation
energy for the three samples with 2 nm thick Si layers. The values
have been scaled to that of defect-free bulk Si. The typical margin
of error for the $S$ parameter is also shown.} \label{fig 1}
\end{figure}

Figure 2 shows how $S$(E) behaves as a function of Si layer
thickness in annealed samples. The features mentioned in the
previous paragraph are present in all samples but their magnitude is
almost negligible in the 1 nm sample. The valley is deepest in the
data from the 4 nm sample but the highest peak is actually found in
the 2 nm sample. The effect of additional annealing is similar in
all samples although the valley at 5 keV deepens considerably more
in the 4 nm sample than in the other two.

\begin{figure}[t]
\centering{\includegraphics[angle=0, width=9cm]{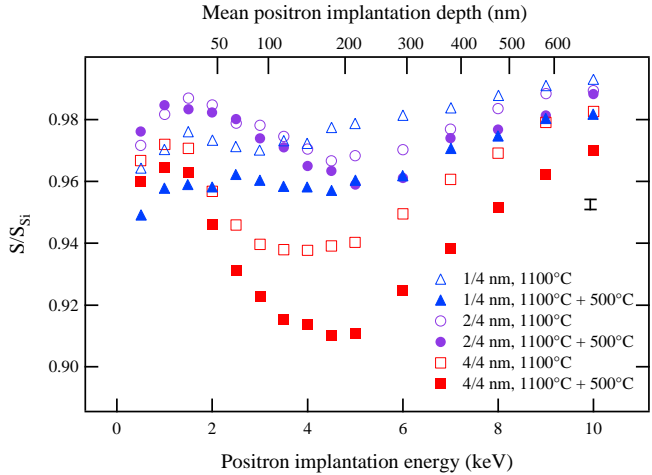}}
\caption{The $S$ parameters as a function of positron implantation
energy for annealed samples with three different Si layer
thicknesses. The values have been scaled to that of defect-free bulk
Si. The typical margin of error for the $S$ parameter is also
shown.} \label{fig 2}
\end{figure}

The photoluminescence results shown in Fig. 3 correlate nicely with
the positron results. The PL signal from the 2 nm sample is the
strongest whereas no photoluminescence at all is observed in the 1
nm sample. The PL peak is positioned at the same wavelength
regardless of sample thickness. This indicates that recombination
does not occur directly over the band gap as the gap width changes
with respect of sample thickness. Thus, the PL signal must come from
defects. Furthermore, no PL was observed from the reference \SiO\
layer either which means that the defects seen in the PL spectra are
located at (\SiO/Si) interfaces and not in the \SiO\ matrix.

\begin{figure}[t]
\centering{\includegraphics[angle=0, width=9cm]{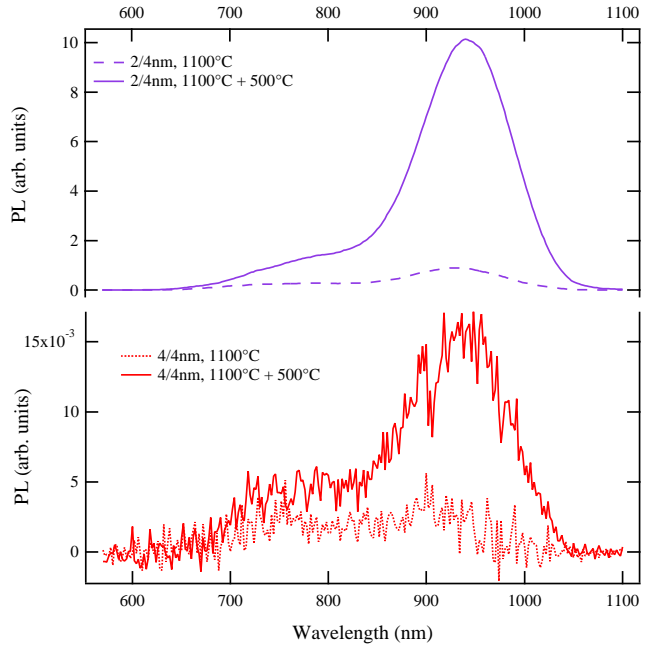}}
\caption{Photoluminescence intensity as a function of wavelength for
the samples with 2 (top panel) and 4 (bottom panel) nm Si layers.
The 1 nm samples are not shown as they exhibit no PL at all.}
\label{fig 3}
\end{figure}

The positron ($S$, $W$) parameters of annealed samples are shown in
Fig. 4. These results clearly reveal the presence of four different
positron states in the samples: the surface (1), two interface
states (2 and 3) and the Si substrate (4). The interface numbered 3
in the figure seems to be the (\SiO/Si bulk) interface reported by
Kauppinen $et~al.$~\cite{Kauppinen1997}. Interface 2, however, has
unique annihilation parameters which most likely correspond to an
unknown defect. Even though the data points coincide with the
SiO$_2$ layer reference point in the 1 and 4 nm samples, preliminary
coincidence Doppler broadening measurements show that the state seen
in these samples does not correspond to SiO$_2$. The signal from the
(\SiO/Si bulk) interface is the strongest in the 4 nm sample whereas
the unknown interfacial state is most clearly seen in the 2 nm
sample.

\begin{figure}[t]
\centering{\includegraphics[angle=0, width=9cm]{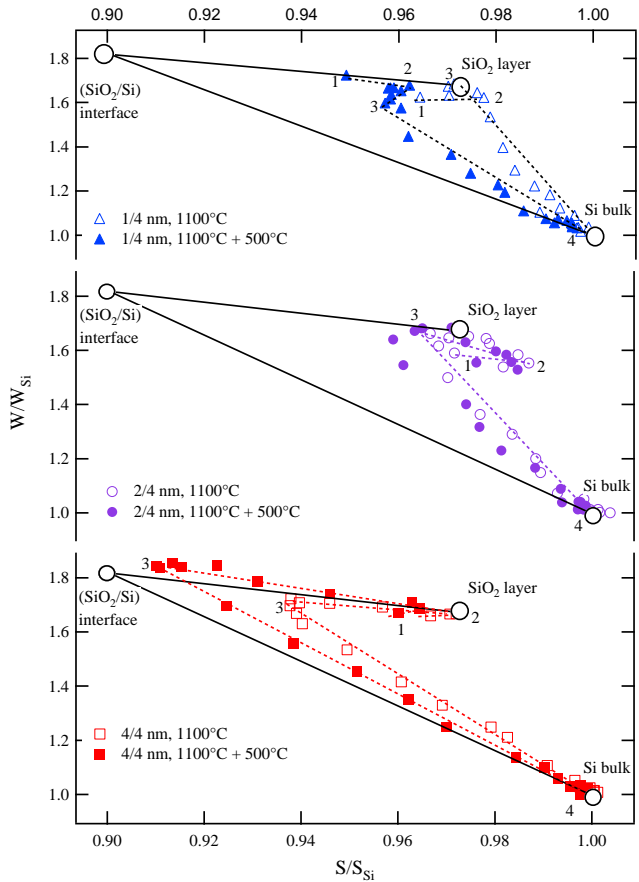}}
\caption{The positron ($S$, $W$) parameters measured in annealed
samples with three different Si layer thicknesses. The values have
been scaled to that of defect-free bulk Si. The bulk Si point, \SiO\
layer point and the (\SiO/Si bulk) interface point from
Ref.~\cite{Kauppinen1997} are also shown. The four annihilation
states: surface (1), unknown interface (2), (\SiO/Si bulk) interface
(3) and Si substrate (4) are marked for each sample. Both the solid
lines connecting the three reference points and the dashed lines
connecting the annihilation states are drawn to guide the eye.}
\label{fig 4}
\end{figure}

The unknown interfacial state seen at positron implantation energies
of roughly 2 keV can be attributed to the interface between the Si
nanoparticles and \SiO\ in the 2 nm and 4 nm samples. The fact that
the positron signal from the nanoparticle interface is the strongest
in the 2 nm samples shows that the nanoparticle formation is
optimized in that sample. The PL results are in agreement with this
observation. At a glance the unknown interfacial state in the 1 nm
samples looks similar to the other samples, especially when the
($S$, $W$) plot of Fig. 4 is considered. However, when looking at
the $S$ parameter data in Fig. 2, it becomes obvious that the
changes in positron diffusion length are much smaller in this sample
than in the other two. In fact, the peak at roughly 2 keV looks more
like a plateau. This indicates that while positrons get trapped by
something also in the 1 nm samples, the trapping state is not the
nanoparticle interface. Most likely another positron-trapping
structure is formed during the annealing in these samples. No PL is
observed from the 1 nm samples so data given by the two techniques
are consistent here as well.

The second annealing step slightly reduces positron trapping into
the nanoparticle interfaces as seen in Figs. 1 and 2. This can be
explained with H atoms filling parts of the open volume at these
interfaces. Photoluminescence from the samples -- as Fig. 3 shows --
increases dramatically following the second annealing step. The
interpretation of these two results is that only a small fraction of
defects present at the nanoparticle interfaces are actually annealed
out during the second annealing step and most are passivated with
the help of hydrogen. Thus, it is evident that the H passivation
step is successful and H ends up at the interfaces.

The signal from the (\SiO/Si bulk) interface getting stronger as the
Si layer thickness increases can be explained with the Makhovian
implantation profile of slow positrons~\cite{Asoka-Kumar1990}. The
profile is highly asymmetric and widens at high positron
implantation energies. This means that in samples where the (\SiO/Si
bulk) interface is located deeper (i.e. the ones with thicker Si
layers) and is probed with higher implantation energies, a larger
fraction of the positrons reaches the interface and annihilates
there.

In conclusion, we have studied novel solar cell materials consisting
of Si nanoparticles embedded within \SiO. The nanoparticles were
formed by annealing Si/\SiO\ multilayer structures in N$_2$ at
1100$^{\circ}$C. The samples were studied with positron annihilation
spectroscopy and photoluminescence. The nanoparticle formation was
shown by both techniques to be the most successful in the sample
with 2 nm thick Si layers. A second annealing step at a lower
temperature (500$^{\circ}$C) involving H was shown to make defects
at or near the nanoparticle interfaces optically passive and thus
enable the use of this material as a solar cell.

Financial support from the Danish Council for Independent
research~\textbar~Technology and Production Sciences (FTP) through
the SERBINA project is acknowledged.

\bibliography{Kirjasto}

\end{document}